\documentclass[preprint,preprintnumbers,
amsmath,amssymb,onecolumn,superscriptaddress]{revtex4}
\usepackage{graphicx} 
\usepackage{dcolumn}  
\usepackage{bm}       
\usepackage{tabularx}
\begin{document}
\preprint{}
\title{Reliability  of  fluctuation-induced transport in a 
Maxwell-demon-type engine} 

\author{Raishma Krishnan}
\email{raishma@gmail.com}
\author{Sanjay Puri}
\email{purijnu@gmail.com}
\affiliation{School of Physical Sciences, Jawaharlal Nehru 
University, New Delhi -- 110067, India.}
\author{ Arun M. Jayannavar}
\affiliation{Institute of Physics, Sachivalaya Marg, 
Bhubaneswar -- 751005, India.}
\email{jayan@iopb.res.in}

\begin{abstract}
We study the transport properties of an overdamped 
Brownian particle which is simultaneously in contact with two thermal baths. 
The first bath is modeled by an additive thermal noise at temperature $T_A$. 
The second bath is associated with 
a multiplicative thermal noise at temperature $T_B$.  
The analytical expressions for the particle velocity and 
diffusion constant are derived for this system, and the 
reliability or coherence of transport is analyzed by means 
of their ratio in terms of a dimensionless P\'{e}clet number. 
We find that the transport is not very coherent, though 
one can get significantly higher currents.
\end{abstract}
\keywords{Brownian motors, Inhomogenous ratchets, Transport coherence.}
\maketitle

\section{Introduction}
The second law of thermodynamics implies that it is not possible to get 
useful work from a single heat bath at constant temperature. The  
perfect time-reversal symmetry (or detailed balance) 
which exists under equilibrium conditions causes the flux in either direction 
to be the same resulting  in zero average flux. 
But advances in this area, especially in the context of 
biological systems, show that it is possible to induce a directed 
motion provided there exist some nonequilibrium 
fluctuations. These are usually provided by unbiased external 
input agents. Devices 
which exploit these fluctuations to generate a unidirectional flow have 
been  termed as ratchets or Brownian motors. 
For a detailed discussion of these systems, see~\cite{rat-gen}. 

The underlying thermal fluctuations are usually 
quantified in terms of temperature. 
Though  these fluctuations cannot be observed directly 
in macroscopic systems, the recent advances in nanotechnology 
and molecular biology allow one to devise molecular systems where 
they play a pivotal role. Thus, rather than trying 
to avoid these inherent fluctuations, they have been utilized for 
constructive purposes: Brownian 
motors serve as the best example of this.

Many different classes of model systems exist in the literature, 
and detailed studies have been done with regard 
to transport and energetics in such systems. 
In our present work, we consider a model of Brownian motor where the 
system has a space-dependent diffusion coefficient $D(x)$. 
The main emphasis in these systems 
is that the potential need not be ratchet-like. Examples 
of such systems arise in semiconductor and superlattice 
structures, molecular motor proteins moving along microtubules, etc.
~\cite{luch,resonance}. 

A space-dependent diffusion coefficient $D(x)$ could arise either 
due to a spatial variation of temperature or friction coefficient, or both. 
For a system having a space-dependent 
temperature [$T(x)$] the system dissipates energy during its evolution 
differently at different places. This implies that the system is 
out of equilibrium. A similar effect arises in a medium with 
a space-dependent friction coefficient [$\eta(x)$] in the presence of an 
external noise~\cite{pareek-amj,mill,jay,land,buttiker}. To be more precise, 
frictional inhomogeneity does 
not generate any current by itself - nonequilibrium fluctuations are 
crucial for transport in these systems.

It has been shown by B\"{u}ttiker ~\cite{buttiker} that, in the 
presence of $D(x)$, the Boltzmann factor $\exp[-V(x)/k_BT]$ 
has to be generalized as $\exp[-\psi(x)]$ where 
\begin{equation}
\psi(x)=-\int ^x dq \frac{v(q)}{D(q)}. \label{1}
\end{equation}  
Here, $v(x)=-\mu dV/dx $ is the drift velocity and $\mu$ is the mobility. 
For a homogeneous system, $D(x)$ is given by Einstein's relation 
$D=\mu k_BT$.  The stability and dynamics 
of nonequilibrium system is greatly 
influenced by the above generalized potential. 
Here, one has to invoke the notion of global stability of states, 
as opposed to local stability which is valid for equilibrium states 
\cite{land}.

In the present work, we consider a Brownian particle in contact with 
two thermal baths, $A$ and $B$. The underlying potential 
$V(x)$ and the friction coefficient $\eta(x)$ are periodic in 
space, and separated by a phase difference 
other than $0$ and $\pi$. We study the quality of transport in such systems.  
Here, the unidirectional current arises 
due to a combination of both $\eta(x)$ and the
fluctuations present in the second thermal bath. 

Millonas~\cite{mill} and Jayannavar~\cite{jay}  have analyzed 
the basic framework for particle transport in such systems. 
Subsequently, Chaudhuri et al.~\cite{banik} have also studied this model 
with modifications in the bath parameters. However, there has been 
no systematic analysis of the dependence of the current on system parameters. 
Further, there is no clear understanding of the coherence of 
transport in these systems. In this paper, we undertake a 
detailed study of both these properties. 

The transport of a Brownian particle is always accompanied by a diffusive 
spread, and the quality of transport is affected by this 
diffusive spread.  This property has been quantified 
via the ratio of current to the diffusion constant and is termed as the 
P\'{e}clet number, Pe. The Brownian 
particle takes a time $\tau=L/v$ to traverse a distance $L$ with a 
velocity $v$ and the diffusive spread of the particle in the same time 
is given by $\langle(\Delta q)^2\rangle=2D\tau $. 
The criterion  to have a reliable 
transport is that $\langle(\Delta q)^2\rangle=2D\tau < L^2$. 
This implies that Pe$=Lv/D > 2$ for coherent transport.
The P\'{e}clet numbers for  some of the models like 
flashing and rocking ratchets show low coherence of transport 
with Pe $\sim 0.2$ and Pe $\sim 0.6$~\cite{low}, respectively. 
Experimental studies on molecular motors showed more reliable transport 
with Pe ranging from 2 - 6~\cite{high}. Our earlier work on 
inhomogeneous ratchets in the presence of a single heat bath but, 
subjected to an external parametric white 
noise fluctuation, showed a coherent transport with Pe 
of the order of $3$~\cite{ijp}. From this and other works~\cite{sch}, 
one concludes that system inhomogeneities may 
enhance the effectiveness of transport, though sensitively 
dependent on physical parameters.

The present paper is organised as follows. Section II gives 
the basic description of the systems, and also the analytical expressions 
for current and diffusion needed to evaluate the P\'{e}clet number. 
In Section III, we present detailed results for the dependence 
of current and quality of transport on system parameters. 
In Section IV, we conclude this paper with a brief summary.  

\section{Model}
We begin with the equation of motion for a Brownian 
particle of mass $m$ and position $x(t)$ coupled simultaneously 
with (a) an additive thermal 
noise bath at  temperature $T_A$, and (b) a multiplicative noise bath 
having a spatially-varying friction coefficient $\eta(x)$ at 
temperature $T_B$. The particle moves in an underlying periodic 
potential $V(x)$.  The equation of motion 
of the Brownian particle is 
given by \cite{mill,jay}
\begin{equation}
 {m\ddot{x}} =-\Gamma(x){\dot{x}}- { {{V^\prime(x)}}} 
+ {\sqrt {{\eta(x)}}}\xi_B(t)+\xi_A(t). \label{leqn}
 \end{equation}
The Gaussian white noises $\xi_A(t)$ 
and  $\xi_B(t)$ are independent, and obey the statistics:
\begin{eqnarray}
  \langle\xi_A(t)\rangle = 0 , \,\,\quad 
\langle\xi_A(t)\xi_A(t^\prime)\rangle &=& 2 \Gamma_Ak_BT_A
\delta(t-t^\prime), \\
\langle\xi_B(t)\rangle=0, \,\,\quad \langle\xi_B(t) \xi_B(t^\prime)\rangle&=& 
2 \Gamma_B k_BT_B\delta(t-t^\prime).
\end{eqnarray} 
The two noises together satisfy the fluctuation-dissipation 
theorem and $\Gamma(x)=\Gamma_A+\Gamma_B\eta(x)$.
 Here $\langle...\rangle$ denotes the ensemble average.
In the present work, we have chosen  $V(x)=V_0(1-\cos x)$ and 
$\eta(x)=\eta_0[1-\lambda \cos(x-\phi)]$. Here $0\,<\,\lambda\,<\, 1$, 
and we set $\lambda=0.9$ throughout for optimum values. The phase 
lag $\phi$ between $V(x)$ and $\eta(x)$ brings in 
asymmetry in the dynamics of the system. This inturn leads to 
an unidirectional current even in the presence of spatially 
periodic (nonratchet-like) potential.

When the friction term dominates  inertia or on time-scales larger 
than the inverse friction coefficient, one can consider 
the overdamped limit of the Langevin equation. This corresponds to 
the adiabatic  elimination of the fast variable (velocity) 
from the equation of  motion by putting $\dot{p}=\ddot{x} = 0$. 
This approach is only applicable for a homogeneous system. 
For an inhomogeneous 
system, Sancho et al.~\cite{sancho} have given the proper prescription 
for the elimination of fast variables.  The resultant overdamped 
Langevin equation for Eq.~(\ref{leqn}) is given by~\cite{pareek-amj}
\begin{equation}
\dot{x} = {- \frac{V^\prime(x)}{\Gamma(x)}}-
\frac{(\sqrt\eta(x))\prime \sqrt\eta(x)}{\Gamma^2(x)}+ 
\frac{\xi_A(t)}{\Gamma(x)}+
\frac{\sqrt\eta(x)} {\Gamma(x)}\xi_B(t).
\end{equation}
The prime on a function denotes differentiation with respect to x.  
Using the van Kampen Lemma~\cite{vankampenlemma} and Novikov's 
theorem~\cite{novikov}, we obtain the corresponding Fokker-Planck 
equation for the probability density $P(x,t)$ of a particle 
to be at point $x$ at time $t$ as follows: 
\begin{eqnarray}
\frac {\partial P(x,t)}{\partial t}&=&  \frac {\partial}{\partial x} 
{\left\{\frac{V^\prime(x)}{\Gamma(x)}  
+\frac{T_B\Gamma_B}{\Gamma^2(x)}
(\sqrt\eta(x))\prime\sqrt\eta(x)\right\}} P +
\frac{T_A\Gamma_A}{\Gamma(x)}\frac{\partial}{\partial x}
\left[\frac{P}{\Gamma(x)}\right] \nonumber \\ 
 && +\, T_B\Gamma_B\frac{\sqrt\eta(x)}{\Gamma(x)}
\frac{\partial}{\partial x}\left[ 
\frac{\sqrt\eta(x)} {\Gamma(x)}P \right].\label{fpeqn}
\end{eqnarray}
When the potential $V(x)$ is positive and unbounded, 
the system evolves to a steady-state distribution $P_s(x)$ characterised 
by zero current. Setting $J(x,t)=0$ in Eq.~(\ref{fpeqn}), we obtain  
\begin{equation}
P_s(x)=N\exp[-\psi(x)],\label{psq}
\end{equation}
where
\begin{equation}
\psi(x)=\int^{x} dq \Big[\frac{V^\prime(q)\Gamma(q)}{T_A\Gamma_A 
+ T_B\Gamma_B\eta(q)} + \frac{(T_B-T_A)}{\Gamma(q)}
\frac{\Gamma_A \Gamma_B \eta\prime(q)}{ T_A\Gamma_A+T_B\Gamma_B\eta(q)}\Big]. 
\label{psi}
\end{equation}
Here, $\psi(x)$ is the dimensionless effective potential, and  
$N$ is the normalization constant. 

For periodic functions $V(x)$ and $\eta(x)$ with periodicity $2 \pi $, 
a finite probability current is obtained. Following Risken~\cite{risken}, 
one readily gets the expression for the total probability 
current $J$ as
 \begin{equation}
 J= \frac{[1-\exp\,({-2 \pi \delta})]}{\int_0^{2\pi} dy \exp\,[-\psi(y)] 
 \int_{y}^{y+2\pi}dx \exp\,\,[\psi(x)]/A(x)}.\label{current}
 \end{equation}
Here, $\delta$ determines the direction of current in the system and 
is given by
\begin{equation}
\delta = \psi(x)-\psi(x+2\pi) \label{delta},
\end{equation} 
%
%
and $A(x)$ in Eq.(~\ref{current}) is given by
\begin{equation}
 A(x)=\frac{\Gamma_A T_A+\Gamma_B T_B \eta(x)}{\Gamma^2(x)}.
\label{aq}
\end{equation}
From Eqs.~(\ref{psi}) -~(\ref{delta}), we can see 
that the system is in equilibrium  when $T_A=T_B$, leading to zero 
net current. But when $T_A \neq T_B$, the system is rendered nonequilibrium 
and one can extract energy at the expense of increased entropy. It can also 
be seen that no current is possible when  either $\Gamma_A$ 
or $\Gamma_B$ is absent. Again, if $\eta(x)$ is independent of 
$x$, there will be no net current flow in the system. When $T_A-T_B$ 
changes sign, the current also changes sign but not the magnitude.

One can also obtain an analytical expression  
for the diffusion constant $D$ by following~\cite{rec,dan-giant} as
\begin{equation}
 D=\frac{\int_{q_0}^{q_0+L}(dx/L)\,A(x)\, 
{[I_+(x)]}^2 I_-(x)}{\left[{\int_{q_0}
 ^{q_0+L}(dx/L)I_+(x)}\right]^3}\label{diffusion}
\end{equation}
 where $I_+(x)$ and $I_-(x)$ are as follows:
\begin{eqnarray}
 I_+(x)&=& \frac{1}{A(x)}\,\,\exp\,[\psi(x)]\,\int_{x-L}^{x} dy 
 \,\,\exp\,[-\,\psi(y)] \, , \label{iplus} \\
 I_-(x)&=& \exp\,[- \psi(x)] \int_{x}^{x+L} dy\,\, 
 \frac{1}{A(y)}\,\exp\,[\psi(y)] \, .\label{iminus}
\end{eqnarray}
Here, $L$ represents the period of the potential and is taken as $2 \pi$. 
From the above equations, the velocity ($v =2\pi J$), 
diffusion constant ($D$) and the P\'{e}clet number (Pe) 
are studied  as a function 
of different physical parameters. All the physical quantities are taken 
in dimensionless form. In particular, $v$ and $D$ are 
normalized by $(V_0/\eta_0 L)$ and $(V_0/\eta_0)$, 
respectively. Throughout our work we have set $V_0 ,\,\eta_0$ and $k_B$ 
to be unity. Similarly, $\Gamma_A, \Gamma_B$ and $T_A, T_B$ are scaled with
respect to $V_0\eta_0$ and $V_0$ respectively. We have used the 
Gauss-Kronord rules for numerical evaluations of the 
integrals involved~\cite{gauskronrod}.

\section{Detailed Results and Discussion}
In  Fig.~\ref{1}, we examine the behaviour of the effective 
potential $\psi(x)$ versus $x$, for two different values of $\phi$. 
The parameter values are  $T_A=1$, $T_B=0.1$, $\Gamma_A=0.1$, 
$\lambda=0.9$ and $\Gamma_B=0.9$. 
For $0 < \phi < \pi$, the 
current is in the negative direction, while for  
$\pi < \phi < 2\pi$, the current is in 
the positive direction. This can be directly inferred from the slope 
of the potentials. The tilt in the effective potential 
identically vanishes when $\phi=0$ or
$\pi$. Hence it is expected that the unidirectional current does not arise for
the case when the phase lag is $0$ or $\pi$. In Fig.~\ref{1}, 
we also study the effect of changing $\Gamma_B$($=0.1$ or $0.9$) on 
$\psi(x)$ for a fixed $\phi=0.3 \pi$ value.
We see that, as $\Gamma_B$ increases, the tilt in effective potential 
increases. We also see from separate plots (not shown here) that 
the tilt is higher for smaller 
$\Gamma_A$ values and also for higher temperature difference 
between the two baths. Hence, we set $T_A=1.0$ and $T_B=0.1$ 
throughout this paper unless specifically mentioned otherwise 
in the captions.


Next, let us study the behaviour of the probability current $J$, 
diffusion constant $D$ and P\'{e}clet number  
Pe for various parameter values 
so as to extract an optimal set of parameters for coherent transport.

\begin{figure}[hbp!]
\begin{center}
\includegraphics*[scale=0.4]{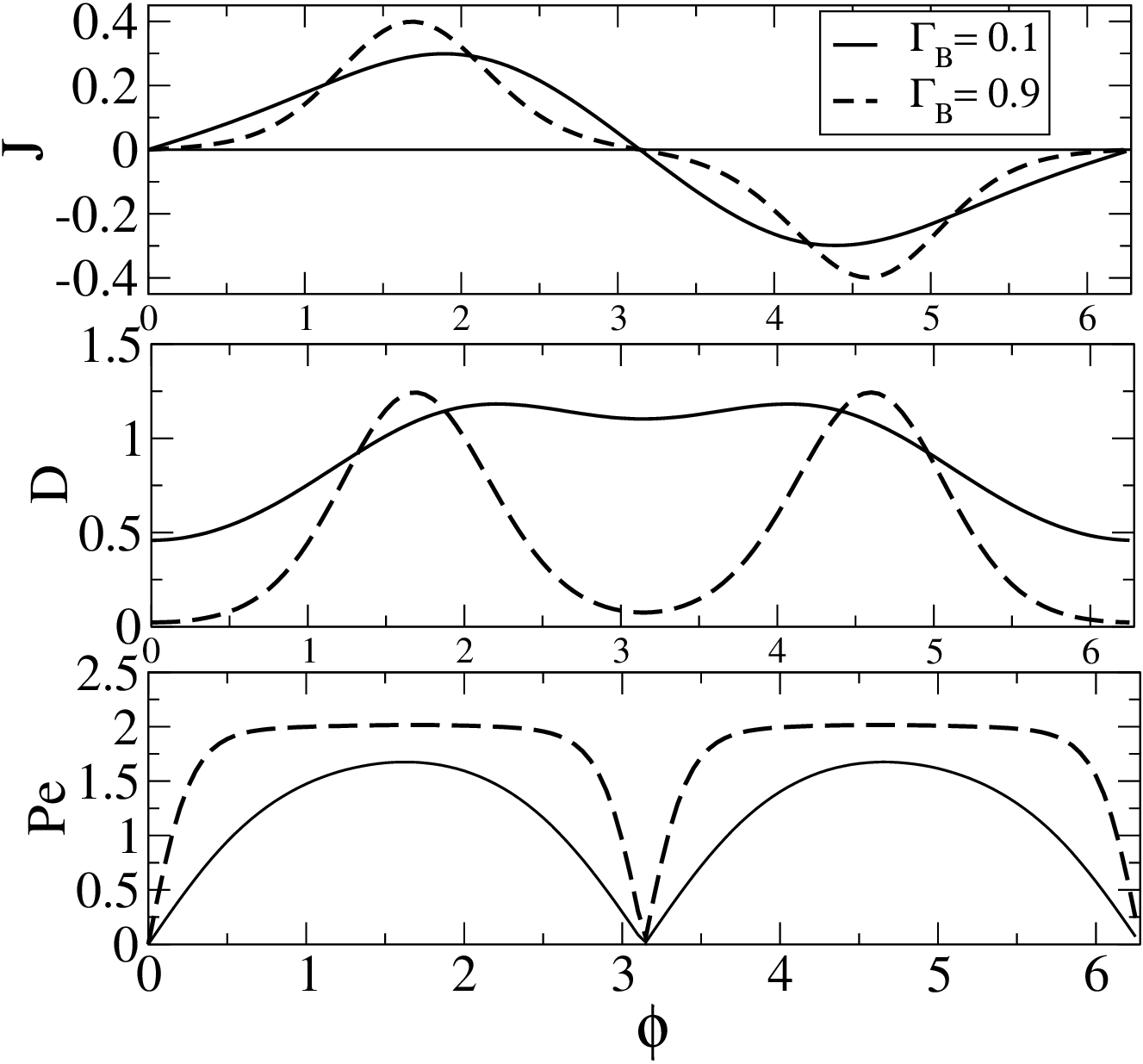}
\caption{Plot of $J$, $D$, Pe vs. $\phi$ for 
$\Gamma_A=0.1$, and $\Gamma_B=0.1\, \mbox{and}\,\, 0.9$. 
For $\Gamma_B=0.9$, the quantities $J$ and $D$ are scaled up by 
a factor of $100$ to make them comparable to the case $\Gamma_B=0.1$.
}\label{2} 
\end{center}
\end{figure}

In Fig.~\ref{2} we plot $J$, $D$ and Pe as a function 
of the phase difference $\phi$ for $\Gamma_A=0.1$, and 
$\Gamma_B=0.1,\,0.9$. The current and 
diffusion constant  varies periodically with $\phi$, as expected. 
We find that an increase in $\Gamma_B$ 
causes a decrease in $J$ and $D$, thereby causing an 
increase in the P$\acute{e}$clet number. We can see from the 
plot of Pe vs. $\phi$ that the transport is marginally coherent ($\sim 2$) 
for a range of $\phi$ - values at the higher $\Gamma_B$ - value.

\begin{figure}[htp!]
\begin{center}
\includegraphics*[scale=0.45]{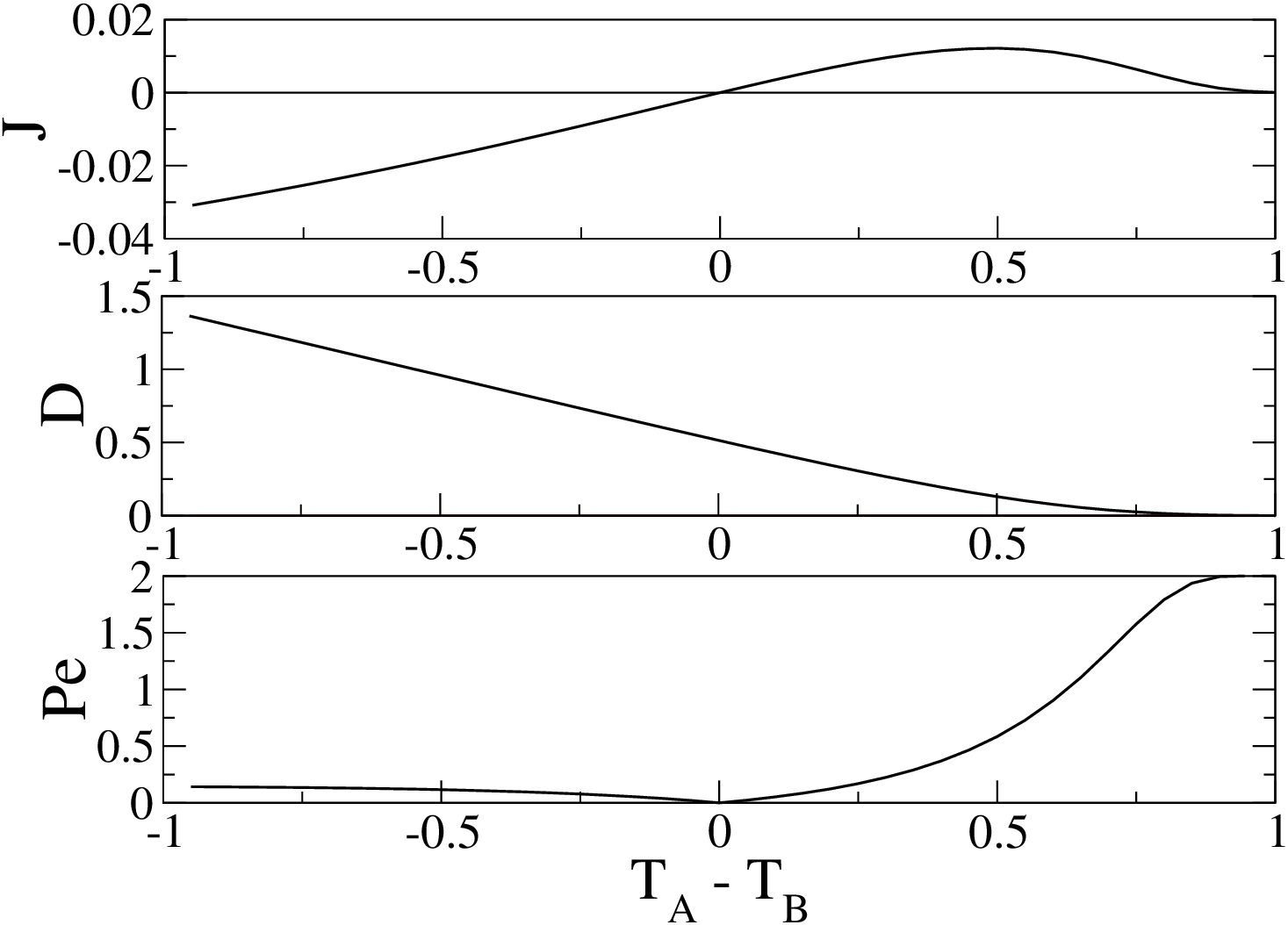}
 \caption{Plot of $J$, $D$ and Pe vs. $ T_A-T_B$ for $\Gamma_A=0.1$, 
$\Gamma_B=0.9$, $\phi=0.3\pi$. We set $T_A=1$ and $T_B$ varies from 
$0$ to $2$.} 
\label{3}
\end{center}
\end{figure}

For a particular value of phase difference ($\phi=0.3\pi$), 
in Fig.~\ref{3} we plot $J$, $D$ and Pe as a function 
of the temperature  difference between the two baths, $T_A-T_B$. 
The other parameter values are  $\Gamma_A=0.1$ and 
$\Gamma_B=0.9$.  Here $T_B$ is varied 
from $0$ to $2$ for a fixed $T_A=1$. As expected, the current 
is zero when the temperatures of the two baths are 
the same. We also observe that the direction of current 
reverses depending upon the temperatures of the two baths $A$ and $B$. 
For $T_B > T_A$, the current is in the negative direction and 
vice versa for $T_B < T_A$. Also, their magnitudes are different in 
both cases.

From Fig.~\ref{3}, we also see that while the current exhibits a peak 
at $T_A-T_B \simeq 0.5$, $D$ decreases with increasing 
$T_A-T_B$. On the other hand, Pe increases with increase in $T_A-T_B$, 
i.e., the transport goes towards the coherent regime.

\begin{figure}[hbp!]
\begin{center}
\includegraphics*[scale=0.45]{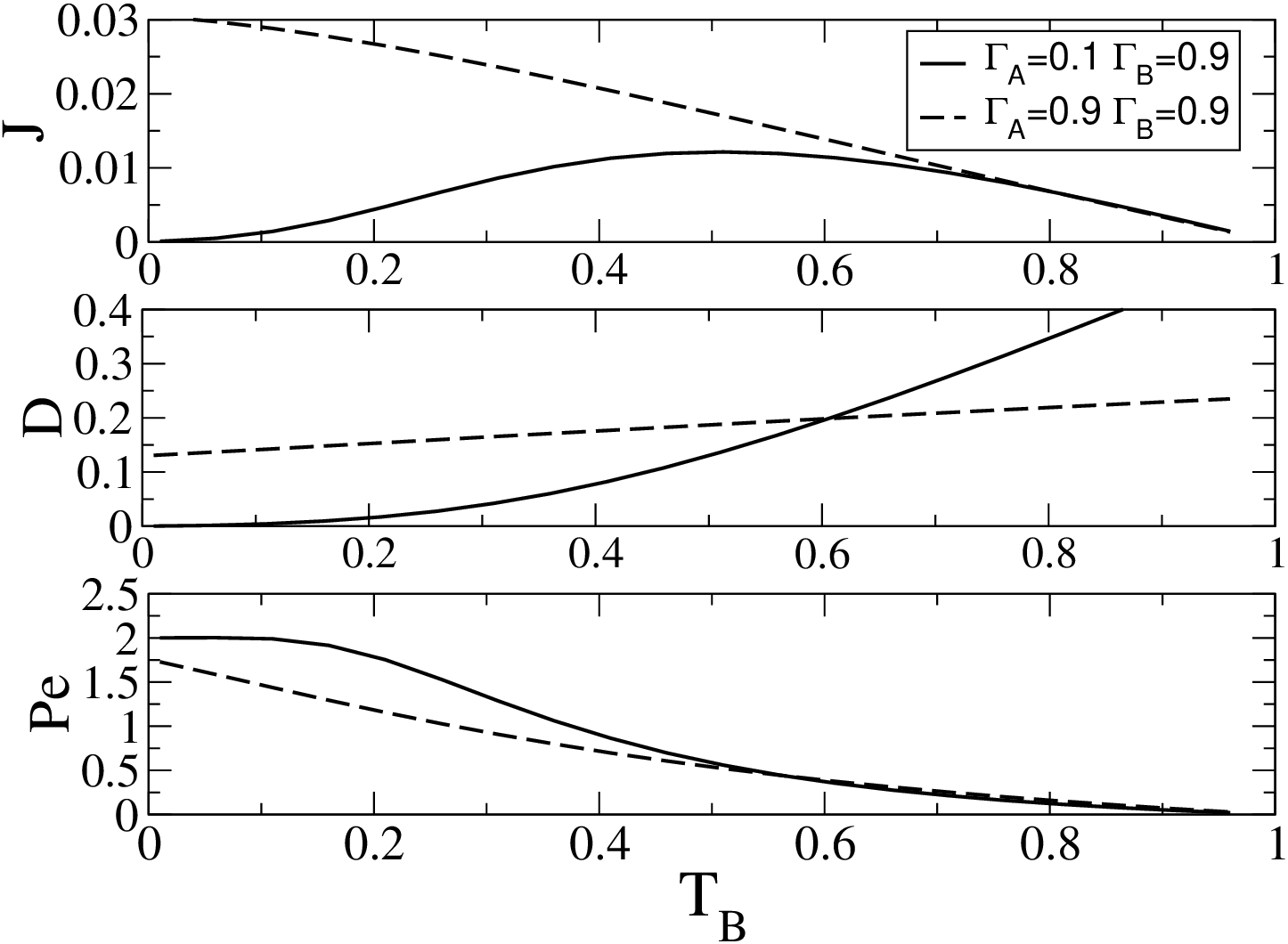}
\caption{Plot of $J$, $D$ and Pe vs. $ T_B$ for 
$\Gamma_A = 0.1\, \mbox{and}\,\,  0.9$, $\Gamma_B=0.9$, 
$\phi=0.3\pi$, and $T_A=1$. } \label{4}
\end{center}
 \end{figure}

In Fig.~\ref{4}, we plot $J$, $D$ and Pe as a 
function of $ T_B$, for $\Gamma_A=0.1$ and $\Gamma_A=0.9$. The other 
parameter values are specified in the figure caption. For $\Gamma_A=0.1$ 
we see  $J \simeq 0$ in the presence of a single heat bath 
$(T_B \simeq 0)$.  As $T_B$ is increased, the current starts to 
build up and shows a peaking behaviour. When the value of 
$T_B \rightarrow T_A$ 
there is no transport as expected. For a particular $\Gamma_B$ value, 
this non-monotonous behaviour is seen only for lower values of 
$\Gamma_A$. When the value of $\Gamma_A$ is increased beyond a 
critical value, there is a non-zero current even 
when the temperature of bath $B$ is zero.

The plot of Pe vs. $T_B$ in Fig.~\ref{4} shows that, as 
$\Gamma_A$ is decreased, the Pe value goes up 
and Pe $\rightarrow 2$ for small $T_B$ when $\Gamma_A=0.1$. 
(We have shown here 
only two representative $\Gamma_A$ - values). We also see that the Pe value 
is more stable for lower $\Gamma_A$ values, i.e., 
Pe $\simeq2$ up to some value of $T_B$, and then decreases with 
further increase in $T_B$. Thus, we conclude that 
$\Gamma_A \leq 0.1$ for higher transport coherence.

\begin{figure}[htp!]
 \begin{center}
\includegraphics*[scale=0.45]{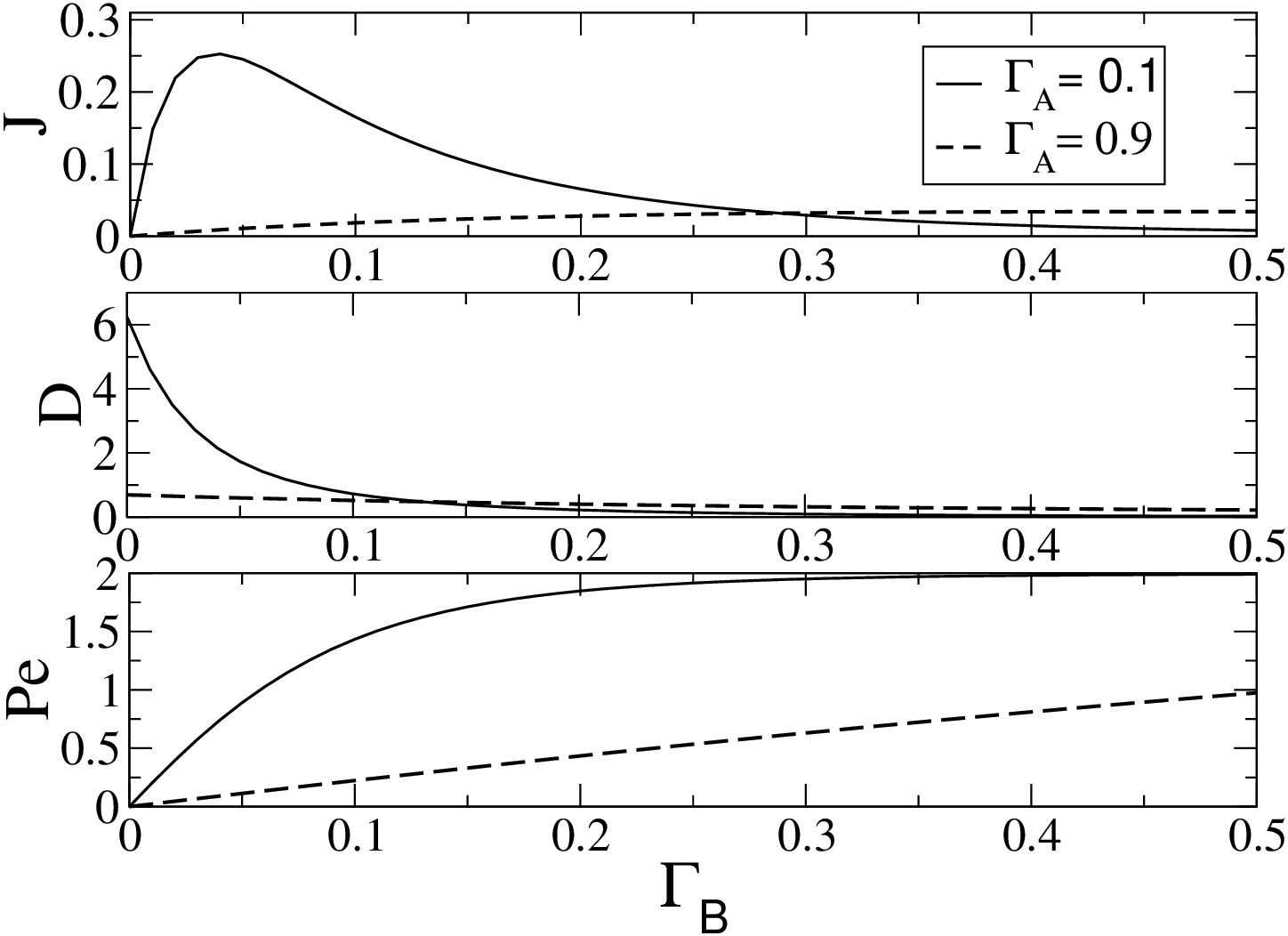}
\caption{Plot of $J$, $D$ and Pe vs. $\Gamma_B$ for $\Gamma_A=0.1\, \mbox{and}
\,\, 0.9$ with $ \phi=0.3 \pi$. The other parameter values are $T_A=1$ 
and $T_B=0.1$.} \label{5}
 \end{center}
 \end{figure}

In Fig.~\ref{5} we plot  $J$, $D$ and Pe vs. $ \Gamma_B$ for 
$\Gamma_A=0.1\,\,\mbox{and}\,\, 0.9$, $\phi=0.3\pi$, 
$T_A=1$ and $T_B=0.1$. In this case, for $\Gamma_A=0.1$, 
the current shows a peak with 
$\Gamma_B$, the diffusion constant decreases with increase 
in $\Gamma_B$ and Pe $\rightarrow 2$ for $\Gamma_B\geq 0.3$. 
On increasing the $\Gamma_A$ value, we see from separate plots 
(not shown here)  that the values of $J$ 
and $D$ are reduced, but the value of Pe is 
almost the same, though  Pe $ >2$ for 
much higher $\Gamma_B$ values, say $\Gamma_B=5$.  
Thus, Fig.~\ref{5} also leads to the conclusion  that a lower 
$\Gamma_A$ - value along with a higher $\Gamma_B$ - value 
aids transport coherence.

\section{Summary and Discussion}
Let us conclude this paper with a brief summary and discussion. 
We have studied the transport coherence 
of an overdamped Brownian particle in the presence of two thermal baths,  
namely $A$ and $B$ with different noise statistics.  One of the baths, $B$,  
is characterized by the presence of a state-dependent 
friction coefficient. The system considered here is analogous to a 
simple model of a Maxwell-demon-type of engine, which 
extracts work out of  the 
nonequilibrium state from thermal fluctuations by rectifying the internal 
fluctuations in the system~\cite{mill,jay}. 
The space-dependent friction coefficient  
in bath $B$ is necessary for generating 
unidirectional current in the absence of a bias. The direction 
of current (but not its magnitude) changes sign when ($T_A-T_B$) 
changes sign. The current identically vanishes when the temperatures 
of both the baths are the same. Also, in the extreme high friction limit, 
the current vanishes as the particle cannot execute Brownian motion.

We have systematically analyzed the behaviour of current, 
diffusion and transport coherence 
as a function of system parameters. The problem of 
coherence of transport has not been addressed in the context of 
these systems.  Our work demonstrates that a combination of lower 
$\Gamma_A$ - value and a higher $\Gamma_B$ - value 
would lead to an optimum transport coherence. Our study of the 
P\'{e}clet number shows that, for most of the 
parameter space, Pe $\leq$ 2. Thus, we conclude from the present study 
that the transport is not coherent though it is possible to get high current.

Our study can be extended in several directions. For example, it would be 
interesting to study the efficiency of energy transduction in this 
Maxwell-demon-type heat engine by using the methods of stochastic 
energetics developed by Sekimoto~\cite{sekimoto}. The fact that 
the steady-state probability distribution $P_s(x)$, Eq.~(\ref{psq}) 
is a nonlocal function of $V(x)$ and $\eta(x)$ itself can lead to 
much interesting physics and would form the basis of our future work. 

\section*{Acknowledgements}
R.K. gratefully acknowledges  DST for financial support through a DST 
Fast-track project [SR/FTP/PS-12/2009].  A.M.J. thanks DST 
for financial support. We also thank Mangal C. Mahato and 
Sourabh Lahiri for useful discussions and help.

\bibliography{draft}

\newpage
\begin{thebibliography}{99}
 
\bibitem{rat-gen} P. H\"{a}nggi and F. Marchesoni, Rev. Mod. Phys. 
{\bf 81}, 387 (2009); P. Reimann, Phys. Rep. {\bf 361}, 57 (2002); 
 H. Linke, Appl. Phys. A 75, 167-352 (2002). 

\bibitem{luch} R. H. Luchsinger, Phys. Rev. {\bf E62}, 272 (2000).  
 
\bibitem{resonance}  M. C. Mahato and A. M. Jayannavar, Resonance 
{\bf 8}, No:7, 33 (2003), M. C. Mahato and A. M. Jayannavar, Resonance 
{\bf 8}, No:9, 4 (2003).

\bibitem{pareek-amj} M. C. Mahato, 
T. P. Pareek and A. M. Jayannavar, Int. J. Mod. Phys. B {\bf 10}, 
3857 (1996);  A. M. Jayannavar and M. C. Mahato, Pramana-J. Phys. 
{\bf 45}, 369 (1995); A. M. Jayannavar, cond-mat 0107080.

\bibitem{mill} M. M. Millonas, Phys. Rev. Lett. {\bf 74}, 10 (1995), 
Phys. Rev. Lett. {\bf 75}, 3027 (1995) . 

\bibitem{jay} A. M. Jayannavar, Phys. Rev. {\bf E53}, 2957 (1996). 

\bibitem {land} R. Landauer, J. Stat. Phys. {\bf 53}, 
 233 (1988). 

\bibitem{buttiker} M. B\"{u}ttiker, Z. Phys. B    {\bf 68},161 (1987).
 
\bibitem{banik} J. R. Chaudhuri, S. Chattopadhyay and S. K. Banik, 
The Journal of Chem. Phys. {\bf 127}, 224508 (2007). 

\bibitem{low} J. A. Freund and L. Schimansky-Geier, Phys. Rev. {\bf E60}, 
 1304 (1999); T. Harms and R. Lipowsky, Phys. Rev. Lett. {\bf 79}, 
 2895 (1997). 

\bibitem{high} M. J. Schnitzer and S. M. Block, Nature (London) 
 {\bf 388}, 386 (1997); K. Visscher, M. J. Schnitzer and S. M. Block, 
 Nature {\bf 400}, 184 (1999).

\bibitem{ijp} R. Krishnan, D. Dan 
and A. M. Jayannavar, Mod. Phys. Lett. B {\bf 19} Nos 19 \& 20, 971 (2005); 
R. Krishnan, D. Dan and A. M. Jayannavar, Physica A {\bf 354}, 
171 (2005);  R. Krishnan, D. Dan and A. M. Jayannavar, 
Ind. J. Phys. {\bf 78}, 747 (2004).

\bibitem{sch} B. Linder and L Schimansky-Geier, Phys. Rev. Lett. {\bf 89}, 
 230602 (2002).

\bibitem{sancho} J. M. Sancho, M. San Miguel and D. Duerr, 
J. Stat. Phys. {\bf 28}, 291 (1982).

\bibitem{vankampenlemma} N. G. van Kampen, Phys. Rep. {\bf C24}, 172 (1976).
\bibitem{novikov} E. A. Novikov, Zh. Eksp. Teor. Fiz. {\bf 47}, 1919 (1964); 
Sov. Phys. JETP {\bf 20}, 1290 (1965).

\bibitem{risken} H. Risken, 
 The Fokker-Planck Equation (Springer Verlag, Berlin, 1984). 

 \bibitem{rec} P. Reimann, C. Van den Broeck, H. Linke, 
 P. H\"{a}nggi, J. M. Rubi and A. Perez-Madrid, Phys. Rev. 
 {\bf E65}, 31104 (2002); P. Reimann, C. Van den Broeck, H. Linke, 
 P. H\"{a}nggi, J. M. Rubi and A. Perez-Madrid, 
 Phys. Rev. Lett. {\bf 87},10602 (2001); B. Linder and 
 L. Schimansky-Geier, Fluct. Noise Lett. {\bf 1}, R25 (2001). 
 
\bibitem{dan-giant} D. Dan and A. M. Jayannavar, 
Phys. Rev. {\bf E66}, 41106 (2002).
\bibitem{gauskronrod} The integrator used here (DCUHRE) is from 
Alan Genz's homepage, 
http://www.math.wsu.edu/math/faculty/genz/homepage. 

\bibitem{d1} D. Dan, A. M. Jayannavar and M. C. Mahato, 
 Int. J. Mod. Phys. B {\bf 14} 1585 (2000); 
D. Dan, M. C. Mahato and A. M. Jayannavar, Phys. Lett. {\bf 258}, 217 (1999).
 
\bibitem{sekimoto} K. Sekimoto, Prog. Theor. Phys. Suppl. {\bf 130}, 
17 (1998); K. Sekimoto, J. Phys. Soc. Japan. {\bf 66}, 1234 (1997).

 \end{thebibliography}
\newpage

\begin{figure}[htbp!]
\begin{center}
\includegraphics*[scale=0.35]{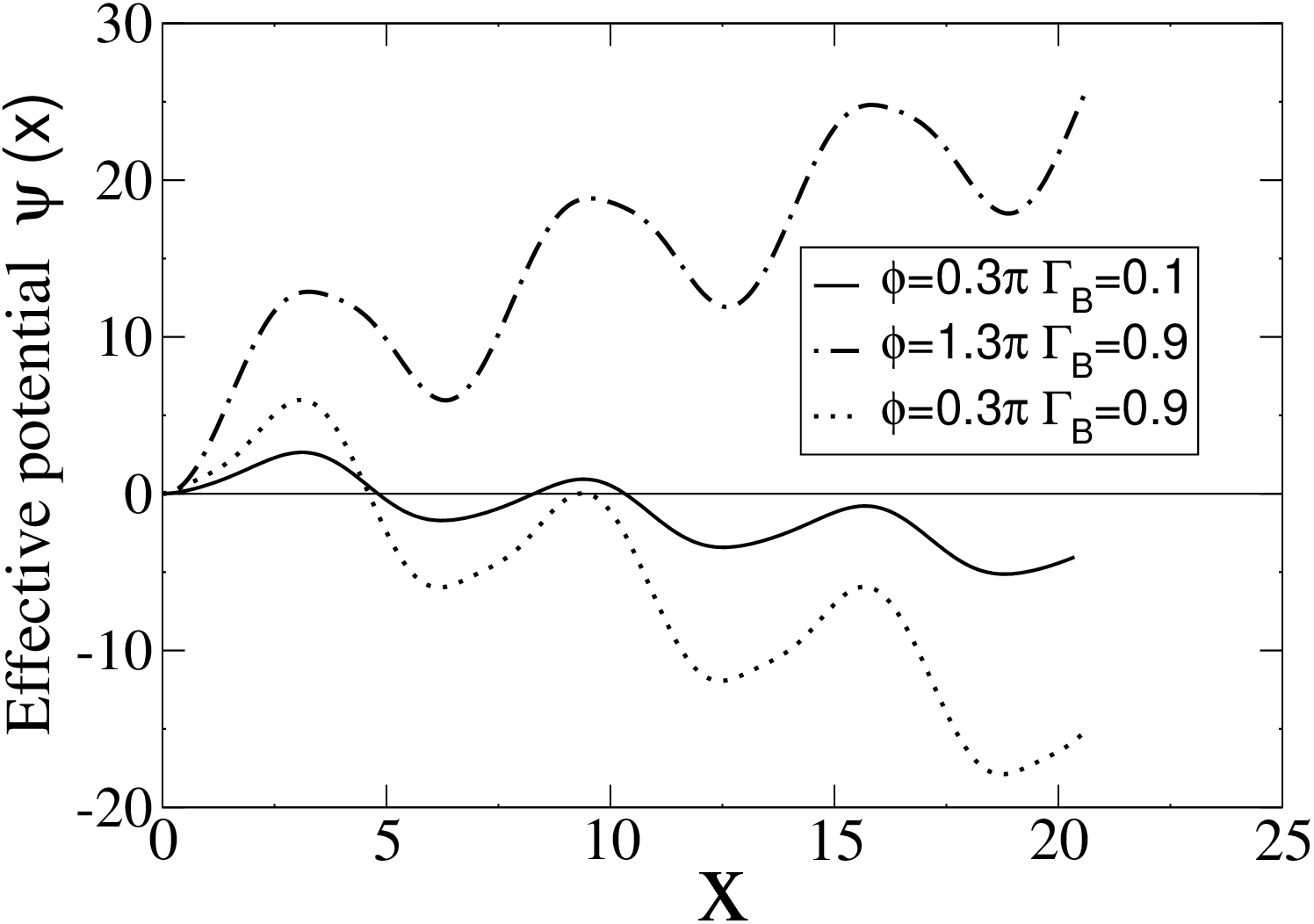}
\caption{Plot of $\psi(x)$ vs. $x$ for the following parameter values. 
(a) $\Gamma_B=0.9$ with $\phi=0.3\pi \,\mbox{and}\,\,1.3\pi$, 
(b) $\Gamma_B=0.1 \, \mbox{and}\,\, 0.9$ with $\phi=0.3\pi$. 
The other parameter values are  $T_A=1$, $T_B=0.1$ 
and $\Gamma_A=0.1$. } \label{1}
\end{center}
\end{figure}
\newpage

\newpage
 \end{document}